# Trust Challenges in Reusing Open Source Software: An Interview-based Initial Study

Javad Ghofrani
University of Lübeck
Lübeck, Germany
javad.ghofrani@gmail.com

Paria Heravi
University of Guilan
Rasht, Iran
paria.heravi@gmail.com

Kambiz A. Babaei
University of Guilan
Rasht, Iran
kambiz.a.babaei@gmail.com

Mohammad D. Soorati
University of Southampton
Southampton, UK
m.soorati@soton.ac.uk

## ABSTRACT
Open source projects play a significant role in software production. Most of the software projects reuse and build upon the existing open source projects and libraries. While reusing is a time and cost saving strategy, some of the key factors are often neglected that create vulnerability in the software system. We look beyond the static code analysis and dependency chain tracing to prevent vulnerabilities at the human factors level. Literature lacks a comprehensive study of the human factors perspective to the issue of trust in reusing open source projects. We performed an interview-based initial study with software developers to get an understanding of the trust issue and limitations among the practitioners. We outline some of the key trust issues in this paper and layout the first steps towards a trustworthy reuse of software.

## 1 INTRODUCTION

Reusing software libraries and open-source projects is an essential part of any software development process [9, 11]. Reusing can increase software quality by decreasing time to market and the risk of encountering unanticipated failure [5, 14]. Despite these advantages, heavy reuse (direct or indirect) creates complex dependencies that are hard for software developers to maintain [4]. As manually maintaining and tracking the required updates for all of the dependencies is a complex task, several automated tools have been recently developed. Apache Ant [15], Apache Maven [12] and Gradle Build Tool [2] are among these tools that facilitate efficient and automated maintenance of dependencies. These tools use a repository to download and attach libraries before building the software projects.

Software-intensive systems usually have complex dependency chains. Any issue in one of the components in the chain can cause failure of the entire software. This is a source of vulnerability that may not reveal itself during the development. The software may run smoothly and pass the testing phase. In 2021, the *log4j* library [1], a logging library that was reused in almost 95% of the java projects at the time, caused a huge damage to the software industry and many companies that were dependent to this library. Within 42 hours of the issue, 800,000 exploitation attempts from unauthorised entities have been registered [2]. Another example of the vulnerability in reused software library is the issue that was reported in Equifax [1] that allowed hackers to steal 147 million users' personal information. Such incidents bring attention to the vulnerability issue that exists in the large dependency chains.

Reusing open source projects faces many challenges [20]. Besides the technical issues, the developers' point of view must be considered. As long as the developers' perspective is not taken into account, we cannot fully utilise the opportunity given by the availability and the variety of the open source projects. Developer's trust has a strong correlation with the quality of the software that they are reusing but the subset of the quality characteristics that have significant role in defining the level of trust is unknown. For instance, a developer might trust a software with high reliability while another may believe that a software with high maintainability is more trustworthy. After creating a model of the trust, we need to measure the difference between an ideal and the existing level of trust and layout a road map to reach the target trust level.

The first step towards the comprehensive outline of the trust issues in reuse is to understand the developers' point of view and measure the awareness among the practitioners. In this study we focus on learning the developers' concerns that limits the trust and present some of the solutions that the developers offered that could allow us to improve the trust level. This paper aims at understanding how well the developers are aware the of the vulnerability of software with heavy reuse and how we can address them. We conducted exploratory interviews with sixteen software developers who were active in the industry for the last 5 to 10 years.

In the next section we go over the related work (Section 2). We then describe our interview study in Section 3 and present the results in Section 4. We outline future work and limitations of our study as well as a conclusion in Section 5.

## 2 RELATED WORK
Related work in this area can be grouped into two categories.

---

[1] https://logging.apache.org/log4j/2.x/security.html

[2] https://www.zdnet.com

## 2.1 Static code analysis

The studies this category mainly look for security issues in repositories and perform static code analysis. They look up the public security announcements and search for related issues in the open source repositories such as *GitHub*. The main issue with these approaches is that the code in repositories may not be well-maintained and the occurrence of issues in outdated projects is not undeniable. For instance, static code analysis was used on open source software repositories to find the correlation between test ratio and test coverage, lines of code, programming language, development methodology and trends, dependency ratio and other metrics to evaluate the vulnerable severity [5–7]. Similarly, Mitropoulos et al. [13] show that bigger projects are more likely to contain security vulnerabilities. Other studies perform dynamic code analysis in addition to static analysis to find a correlation between the known security vulnerabilities and the source code[18]. They also propose a tool that help in mitigating vulnerability. In an empirical study, Kula et al. [10] analysed GitHub Projects to find security vulnerabilities in their dependencies and observe that those library dependencies will not be actively updated. Prana et al. [19] scanned some of the Java, Python and Ruby open source projects to find vulnerable dependencies. They used a novel extracting method and analysed registered metadata in repositories such as commits and fixes that were used for the maintenance of the projects. Wermke et al. [21] conducted interviews with 27 people who were involved in the maintenance of open source projects and analysesd their practices for security and trust.

## 2.2 Dependency chain tracing in software ecosystems

The second set of studies focus on tracing public security announcements in dependency chain of the software ecosystems (e.g., Maven, PyPI, RubyGems, nmp). These methods are more effective compared to static code analysis as they reveal the vulnerability in dependency chains that can lead to security issues in the fully-developed projects. The argue that due to the high dependency density in software ecosystem packages, any simple security issue can affect a large number of programs. For instance, Hejderup et al. [8] propose a method that analyses dependency graphs in the code and also traces the dependencies through the repositories of reused packages. Another study analysed the evolution of *npm* open source packages over the course of 6 years to find correlations between the discovered vulnerabilities and the software packages[3]. Their study suggests that the package maintainer should act more actively in updating the fixes for security vulnerabilities and informing the developers about the reported issues and updates. Zimmermann et al. [22] report that npm packages from npm ecosystem suffer from unmaintained packages, even years after the publicly announced security issues. Their proposed mitigation method focuses on training the maintainers and also performing security tests on the npm repository before updating a new library. Pashchenko et al. [16] report that not all vulnerabilities may cause sever security issues in the industrial projects. They also developed a methodology to prioritise the reported vulnerabilities and available fixes to help practitioners to focus on most important issues and fix them [17] .

All of these approaches take the perspective of the developers and the maintainers of the original code rather than the software developers that reuse these projects. We take the latter approach to investigate the human factors and trust issues in reusing open source projects.

## 3 SURVEY

### 3.1 Preparation

We designed an interview questionnaire after a series of brainstorming sessions with the authors and a group of researchers at the university of Luebeck. We have started the pilot interview with five software developers. Three of them were in the same level of knowledge and experience as our main interviewees. 2 of them were junior software engineering with less than 2 years of experience. First, we measured the time required for conducting the interview. Then, we considered that the all of test interviewees have similar understanding and interpretation of the questions. This step helped us to remove double meaning of the questions and prevent them in the main interviews. After three iterations, the questions were finalised and we started the interviews. In order to produce robust and reliable results, we selected mid-level developers with at least 5 years of experience. We have searched in social networks, especially LinkedIn, for the senior developers. We checked their timeline to make sure that they had at least 5 years of experience in software development. Our goal was to have enough diversity in field of work (e.g. back-end development, front-end development, hardware programming, windows applications, etc.). We contacted more than 100 people from different work domains.

### 3.2 Structure

The questionnaire consists of three sections. The first part (Section A) is designed to collect the introductory information including the experience and the field/domain of expertise. The second part (Section B) asks the interviewees for the activities and guidelines with regards to reusing open source software components in their current organisations. The final part (Section C) of the questionnaire collects their opinion regarding the challenges and the solutions for mitigating the security issues of reuse. At the beginning of the interview, the participants were briefed about the data collected during the interviews and the anonymity of the published results. The participants were informed about the objectives this research. In order to prevent the bias, the participants were not aware of the hypothesis and the general purpose of this study until after the interview sessions. Table 1 shows the list of questions asked during the semi-structured interviews.

## 4 RESULTS

We conducted 16 interviews in total. 2 interviews were offline and 14 other interviewees were online and transcribed. All interviewees and interviewers working in different companies. 14 interviewees work in Iran while 1 works in UK and 1 works in Germany. We transcribed the interviews and put the answers to each question in one table. Then we coded the transcribed text and assigned keywords to each answer. We created a clustering of keywords and grouped similar keywords together. Then, we mapped the answers to their corresponding clusters. The limitations and solutions arose

Table 1: List of questions used for our semi-structured interviews.

| | | | Question (main question, follow-up question) |
|---|---|---|---|
| Introductory questions | MQ | | **Part I:** Describe your organization and your role. How many years of experience do you have? What are the activities that you are mainly involved in? What is the main software development methodology in your organization? |
| | | FQ | What is the main sector of your organization? What are the services and products that your organization provides? |
| | | FQ | Is your organization independent or is it a subsidiary of another company? Does your company have active partnership with other companies? If yes, what are the type of this partnership? |
| | | | **Optional:** Was there a major technological or structural change in your organization during the last three years? If yes, describe the changes? |
| | | FQ | **Part II:** How do you define reuse of libraries and software modules? In your organization, how much do you reuse other modules or libraries (Low, High)? Do you follow a certain guideline or standard for reuse |
| Key questions | MQ | | Open Source Projects : How much do you use open source projects (estimated percentage)? |
| | | FQ | In general, what makes you use open source projects? What are the key factors that you consider in choosing an open source project to reuse? |
| | MQ | | Do you think that the scale of reuse has changed since you joined your organization? If yes, how much (Scale one to five) |
| | | FQ | After the change: How was your experience? How did it affect your activities and tasks? What was improved in your activities? What were the new challenges? |
| | MQ | | Do you pay attention to the security concerns of reusing open source projects? If yes, how do you deal with the security concerns? If no, describe the reason for the negligence. |
| Future issues | MQ | | **Part III:** (We have asked for your understanding and knowledge so far, the rest of the questions are focused on your opinion.) Aside from the organization's perspective, in your opinion, what are the key challenges, risks and common mistakes in reusing open source projects? |
| | | FQ | What are the solutions for mitigating these issues? Are there any challenges in implementing those solutions? If yes, describe them. |

from discussions between the co-authors based on the contents of each cluster.

Table 2 shows the distribution of the interviewees over their domain of work. Among them we have 4 Front-end Developers, 2 lecturers, 2 AI Researchers, 1 CTO, 1 DevOps Engineer, 1 Network security researcher, 1 Hardware developer, 3 Mobile app developers and 4 Back-end developers.

Table 3 lists the working experience of the interviewees in the company where they currently work. Most of them have between 1-3 years of working experience, one interviewee has more than 15 years of experience, and two interviewees have between 10-15 of experience. Note that, our interviewees have more than 5 years of work experience. The values listed in Table 3 show the experience of the interviewees in their current organisations.

| Participant's Role | Number of Answers |
|---|---|
| Front-end developer | 4 |
| Back-end developer | 4 |
| Mobile app developer | 3 |
| AI researcher | 2 |
| Lecturer | 2 |
| DevOps engineer | 1 |
| Network security researcher | 1 |
| Hardware developer | 1 |
| CTO | 1 |

Table 2: Domain of the work of interviewees

| Years of work in current organisation | Answers |
|---|---|
| between 1 and 3 years | 4 |
| between 3 and 5 years | 3 |
| between 5 and 10 years | 3 |
| between 10 and 15 years | 1 |
| more than 15 years | 2 |
| not mentioned | 3 |

**Table 3: How many years of experience do the interviewees have with their current organisation**

| Frequency of reusing open source projects | Answers |
|---|---|
| more than 70% | 8 |
| between 30% and 70% | 3 |
| less than 30% | 5 |

**Table 4: Frequency of reusing open source projects**

Table 4 lists the number of interviewees who said how much they reuse open source modules in their projects. It means, we asked how often the interviewees reuse code in their projects. With "less than 30%" meant that none of their projects was based on reuse of open-source projects and they do almost everything from scratch, while "more than 70%" meant that most of the time, if they should develop new functionalities or software, they prefer existing package and software repository to reuse. Half of them (8 from 16) use open source projects more than 70% in their projects while 5 of them reuse open source projects less than 30% in their projects.

We extracted the key expressions from the interviewees. When asked about the considered factors in choosing a library or open source project for reuse, 5 participants mentioned "Relevance to current project", 2 mentioned that they check if there is a decent documentation available. 4 of them mentioned that they consider the number of downloads. 2 of them check if there is enough updated tests in the project. 3 consider the quality factors of the code such as clean code, availability and extendibility of the projects. 5 of them mentioned the size of the developer community and 5 check the lifeline of the projects. 3 participants mentioned that the user communities and their reviews about the project are important to them. 5 look into the recent commits and 2 check the security issues. 3 keywords mention the ranking stars from GitHub.

For the question regarding the developers' consideration for the security of reused modules, 12 stated 'Yes' and 4 said 'No'. Only 3 of the interviewees check the security and vulnerability of the reused codes themselves. 5 of them have an extra security team that decide and define the policies and the guidelines for allowing or forbidding the reuse of libraries and dependencies. 5 have no special mechanism to control the trust and the security of the reused code.

The participants were asked about their opinion regarding the key challenges, risks and common mistakes in reusing open source projects and what their suggested solutions for overcoming those challenges are.

### 4.1 Identified Limitations

Figure 1 shows the main limitation for trust in reusing open source software that extracted from our interviews. There are 5 key issues as listed below.

*4.1.1 Lack of continuous support.* **Limitation:** The users were concerned that the open source project developers may abandon the project and leave the project without continuous support, updates, commits and bug fixes. Even big projects can be stopped after a while since the main developer or development team leave the community or the project. This issue is mentioned by 5 interviewees.

**Suggested solutions:** Sponsorship, long-term participation on donation and collective payment methods should be established for motivating and supporting the developers to continue their work (mentioned by 3). Other suggestion was to solve this challenge by checking which developer community/development team are driving the package/library. It is also recommended to (re)use the code and libraries only from strongly supported communities like *Eclipse Foundation* or *Apache Software Foundation*.

*4.1.2 Maintenance Cost.* **Limitation:** Open source projects cannot meet the full requirements and objectives of the users' projects. Therefore, reusing open-source projects requires additional effort to tailor, extend or integrate the existing project to meet the end-users' goals and requirements. There is always the risk that the integration of open-source projects may lead to additional integration effort rather than saving the time or improving the quality as it is expected from reusing. Some new security vulnerabilities can be introduced during the integration and manipulation of the reused code. Furthermore, tracing the published updates and the integration over time can have more overhead compared to the modules that are entirely developed from scratch. One participant mentioned that the unnecessary parts of reused software code should be removed. Otherwise it can cause maintenance problems and introduce additional costs in the future. Another interviewee stated that in the case of obsolesce of reused dependencies, migration can add cost and may lead to new problems. Replacing new libraries require a tedious effort of extracting the dependencies to the old library and maintaining the software after such change can be very costly.

**Suggested solutions:** 2 Participants believe that reducing branching of the open source projects and intention to use open source packages and libraries can help solve the issue. One interviewee suggested to reuse the idea instead of reusing the code. The participant explained that rather than direct reuse of the open source projects, developers should get inspired and study their methods for to develop their own projects. Another suggestion was to perform code reviews before reusing to reduce cost at the later phases.

*4.1.3 Lack of Alternatives.* **Limitation:** Developers may reuse because there is a pressure from product owners and project managers or they do not have enough competence or skills to develop a project. The reuse in this case is not due to their informed decision with options but because of there is no other viable alternative. 6 interviewees raised this issue.

**Suggested solutions:** An interviewee suggested to use scanning tools before reusing the packages or open source projects. Another interviewee mentioned that the developers should consult with their team and ask for other experiences and opinion before reusing.

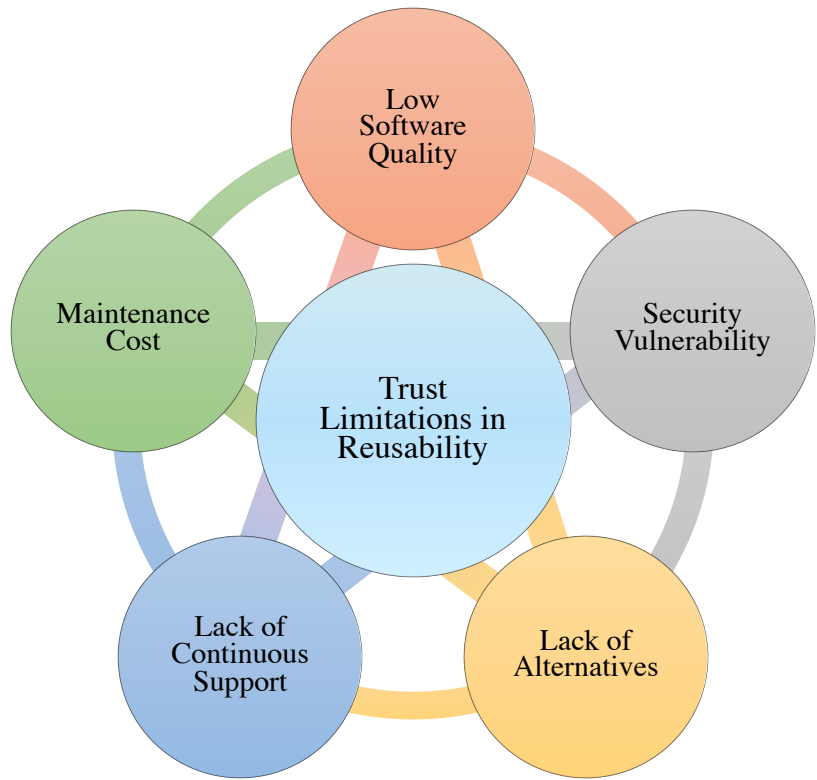

Figure 1: Interconnected limitations of trust in reusing open source software from the point of view of developers.

*4.1.4 Security Vulnerabilities.* **Limitation:** 5 interviewees raised this issue as a challenge in reusing open source libraries. Based on their input, known and unknown bugs and issues in both libraries and software codes can lead to security issues in the end product. Furthermore, some projects are so huge that the developers forget to remove unnecessary parts from their own project after reuse which reduces the performance of the software and opens new potential vulnerabilities.

**Suggested solutions:** an interviewee suggested isolation policies for reused packages and libraries. The participant believed that there should be limitations in the architecture of the projects. This way, if some exploits try to use the security vulnerabilities of reused packages, the damage can be less harmful. However, the trade-off between the effort for isolation and the benefits of reuse should be considered. One other interviewee suggested periodic security reviews which includes automatic and semi-automatic usage of scanner tools and checking updates, fixes and dependencies. Another interviewee advised against directly using the last version of packages and libraries after release. According to his/her suggestion, using penultimate version of packages and libraries can prevent security issues.

*4.1.5 Low Software Quality.* **Limitations:** One participant said that although software reuse can increase the quality of the software product, it should be applied in a reasonable way. A wrong way of reuse can cause performance issues. Lack of documentation is one of the reasons that makes it hard to search and understand the reusable code and packages. Furthermore, low quality of reused code makes it hard to understand, extend and maintain such projects.

**Suggested solutions:** The interviewees mainly believed that we should consider the match level between requirements of the project and the existing libraries. They also suggested to perform a quality estimation based on well-known metrics (updates, test, comments of user community, etc.) before deciding to reuse a library or a package.

## 5 CONCLUSION AND DISCUSSION

This paper focuses on the issue of vulnerabilities in open source software packages from the perspective of the developers that reuse them. We reflected the state of the practice and studied the developers' awareness and trust. We interviewed 16 developers from different domains and asked for their point of view on the main risks of reusing open source software. We asked them about the risks and their proposed solutions to mitigate the identified risks. We collected and analysed the results and presented a trust limitation factors in this paper. We identified five key limitations in reuse that are lack of continuous support, maintenance cost, low software quality, security vulnerability and lack of alternatives.

The results of interview show that the developers are well-aware of the risks and have a justified level of trust in third party open source projects and libraries. However, the proposed solutions mainly lack a proposition to use automated tools or systematic methods. The suggested solutions are mostly based on the developers' experience rather than any existing framework or tools. This implies that the developers see themselves responsible for any issues and this is a challenge that needs to be handled manually. Despite our classification of concerns, it needs to be noted that the limitations are interconnected. For instance, lack of continuous support will gradually lead to security vulnerability and a project that has many vulnerabilities may cost so much that the software will not be maintained anymore.

## 5.1 Threat to Validity

The results in this paper are a summary of observations and should not be used as a proof, proposal or guideline. This paper presented a sample of viewpoints in order to put forward an initial study. However, there are some points that should be discussed regarding the validity of this paper.

*External Validity:* Although 14 out of 16 interviewees work in Iran, we could not find any evidence that the open source culture in Iran differed from other countries around the world. However, this is still a threat to validity that we could not exclude in this research. Thus, more research in a bigger scale is required, which we planned for future work. Furthermore, regarding population validity, we do not believe that the results in this paper are representative of the community of the developers. However, we tried to collect various opinions from different domains of work. Still there are limitations in terms of number, location, domains, experience and gender which should be considered when this work is reused or extended. A last point about the study's external validity regards language. We excluded misunderstandings and misinterpretations of interview questions by conducting the interviews in the native language of interviewees.

*Internal validity.* In order to prevent errors and mistakes in the classification of results, all authors are involved in the classification process. We did not translate the answers to English to exclude misunderstandings, but the assigned keywords are decided trough discussion by the authors.

## 5.2 Outlook

There are plenty of opportunities to build on our study. One direction for future work is to perform a bigger and comprehensive interviews that covers many working domains and includes diversity. We also encourage future studies that focus on evaluating the effect of preventive methods on solving security issues in reusing open source software.

## 6 ACKNOWLEDGMENTS

We would like to thank the interviewees who provided shared their knowledge and opinion with us. The participants may not necessarily agree with our interpretation and conclusion but their inputs were essential for this manuscript. We also acknowledge funding from University of Luebeck, Germany.